\documentclass[fleqn,twoside]{article}
\usepackage{espcrc2}
\usepackage{amsmath}
\usepackage{amssymb}
\usepackage{graphicx}
\usepackage{subfigure}
\usepackage{cite,mcite}
\usepackage{xspace}

\def\EeV{\ifmmode {\mathrm{\ Ee\kern -0.1em V}}\else
                   \textrm{Ee\kern -0.1em V}\fi\xspace}%
\def\PeV{\ifmmode {\mathrm{\ Pe\kern -0.1em V}}\else
                   \textrm{Pe\kern -0.1em V}\fi\xspace}%
\def\TeV{\ifmmode {\mathrm{\ Te\kern -0.1em V}}\else
                   \textrm{Te\kern -0.1em V}\fi\xspace}%
\def\GeV{\ifmmode {\mathrm{\ Ge\kern -0.1em V}}\else
                   \textrm{Ge\kern -0.1em V}\fi\xspace}%
\def\MeV{\ifmmode {\mathrm{\ Me\kern -0.1em V}}\else
                   \textrm{Me\kern -0.1em V}\fi\xspace}%
\def\eV{\ifmmode {\mathrm{\ e\kern -0.1em V}}\else
                   \textrm{e\kern -0.1em V}\fi\xspace}%
\newcommand{\Xmax}{\ensuremath X_\mathrm{max}\xspace}
\newcommand{\meanXmax}{\ensuremath \langle X_\mathrm{max}\rangle\xspace}
\newcommand{\gcm}{g/cm$^2$\xspace}

\title{Composition Studies with the Pierre Auger Observatory}

\author{M.\ Unger
       \address[KIT]{
        Karlsruher Institut f\"ur Technologie (KIT)\\
        Postfach 3640, D-76021 Karlsruhe, Germany}
        for the Pierre Auger Collaboration
        \address{Observatorio Pierre Auger\\
        Av. San Martin Norte 304, 5613 Malarg\"ue, Argentinia}
}%
       
\begin{document}

\begin{abstract}
 We report on studies of the composition of ultra high energy
 cosmic rays with the Pierre Auger Observatory.
 The detection of longitudinal air shower profiles with
 the fluorescence detector is described and the measurement
 of the average shower maximum as a  
 function of energy is presented. Furthermore, mass sensitive
 parameters that can be obtained from the observatory's surface
 detector data are discussed.
\end{abstract}

\maketitle
%
%
\section{Introduction}
The Pierre Auger Observatory is a facility to study the origin of
ultra high energy cosmic rays by a combined measurement of their flux,
anisotropy and mass composition.

The southern site of the observatory is located near the town of
Malarg\"ue in Argentina and its construction has been completed at the
end of 2008. Even during its assembly, data have been collected in a
stable manner since January 2004.  The lateral densities of charged
air shower particles at ground level are measured with the surface
detector (SD), an array of 1600 water Cherenkov detectors distributed on an
area of 3000 km$^2$. In addition, the longitudinal air shower development
can be observed with the fluorescence detector (FD), consisting
of 24 wide-angle Schmidt telescopes that overlook the atmospheric
volume above the surface detector.

The mass composition of cosmic rays above 10$^{17}$~\eV is a key
observable to discriminate between different models put forward to
explain the softening of the cosmic ray energy spectrum at energies
between 10$^{18}$ and 10$^{19}$~\eV (the so called 'ankle',
see~\cite{Berezinsky:2007wf} for a recent review). Moreover, a
measurement of the ultra high energy cosmic ray composition is
essential to understand the nature of the flux suppression observed
above~5$\cdot10^{19}$~\eV~\cite{bib:gzk}.

In this article we will explain, how the cosmic ray composition can
be inferred from air shower measurements with the Pierre Auger
Observatory.  An estimate of the primary composition from the
measurement of the longitudinal air shower profiles with the
fluorescence detector will be given in the next section, followed by
the description of the mass sensitive observables of the surface
detector.
%
%
\section{Fluorescence Detector Measurements}
Within the Heitler model~\cite{bib:heitlerModel} of cosmic ray
induced particle cascades, the maximum number of electromagnetic
particles of a shower with energy $E$ and mass $A$ 
is reached at a depth of
\begin{equation}
  \meanXmax = \alpha + X_0\left(\ln E - \langle\ln A\rangle\right)\,
\end{equation}
where $X_0$ denotes the electromagnetic radiation length in air.
The details of the hadronic interactions feeding the shower
are subsumed into the parameter
$
  \alpha = \lambda\ln 2 - X_0 \ln\left(3N_\mathrm{ch}E_\mathrm{crit}\right),
$
where $\lambda$ denotes the hadronic interaction length, $N_\mathrm{ch}$
is the charged particle multiplicity and $E_\mathrm{crit}$ the 
energy threshold, below which ionization becomes more important
than radiative processes.
The change of $\Xmax$ per logarithm
of energy is called elongation 
rate~\cite{bib:elongationRate}:
\begin{equation}
   D=\frac{\mathrm{d}\alpha}{\mathrm{d}\ln E}+X_0\left(1-\frac{\mathrm{d}\langle\ln A\rangle}{\mathrm{d}\ln E}\right)\;.
\end{equation} 
Thus, given a prediction of $\alpha$ and 
$\frac{\mathrm{d}\alpha}{\mathrm{d}\ln E}$ from air shower simulations,
the primary mass composition
can be estimated from the average shower maximum $\meanXmax$ and its relative
change with energy from the elongation rate. In the following we will
describe the measurement of these two quantities with the Auger fluorescence
detector.
\begin{figure}[t!]
  \centering
  \includegraphics[width=0.8\linewidth]
   {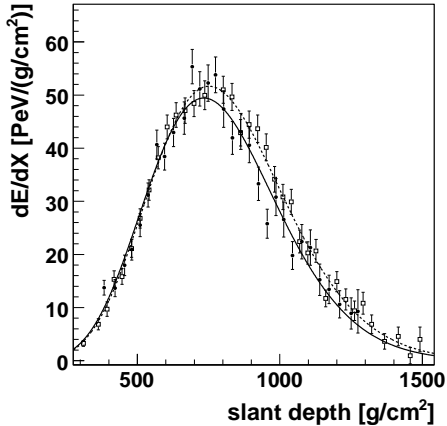}
  \caption{Example of a measured longitudinal profile,
           reconstructed independently by two
           fluorescence detectors: Los Leones (black dots, E=31$\pm$2 \EeV, 
           $\Xmax$=733$\pm$18~\gcm) and Los Morados (open squares, E=33$\pm$2 \EeV,
           $\Xmax$=751$\pm$16~\gcm).}
  \label{fig_Profile}
  \vspace*{-0.3cm}
\end{figure}
\begin{figure}[t!]
  \centering
  \includegraphics[width=0.66\linewidth]
   {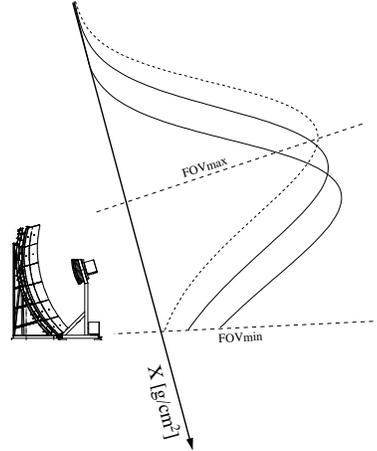}
  \caption{Sketch of the field of view of a fluorescence telescope and
           three longitudinal profiles. Showers with their maximum above
           FOVmax or below FOVmin are not selected.}
  \label{fig_fov}
  \vspace*{-0.3cm}
\end{figure}
\subsection{Event Reconstruction}
In the FD composition analysis only events recorded in hybrid mode
\cite{bib:hybrid} are considered, i.e.\ the shower development must
have been measured by the fluorescence detector and at least one
coincident SD station is required to provide a ground level time that
is needed for a precise determination of the shower direction.  The
longitudinal energy deposit profile is
reconstructed~\cite{bib:profileRec} from the light recorded by the FD
using the fluorescence and Cherenkov yields
from~\cite{bib:fluorescence} and~\cite{bib:cherenkov}. With the help
of data from the observatory's atmospheric monitoring
devices~\cite{bib:augeratmo} the attenuation of the light on its way
from the shower to the detector is corrected for and the measured
light emission heights are converted to atmospheric depth.
Fig.~\ref{fig_Profile} shows an example of a reconstructed
longitudinal profile of an air shower, that triggered two of the four
Auger fluorescence detectors.  The data is fitted with a
Gaisser-Hillas function~\cite{bib:ghFunc} to extrapolate the profile
to depths outside the field of view of the fluorescence telescopes.
The integral over this function gives the calorimetric energy of the
shower and the depth where it is maximal is at $\Xmax$.  From showers
that were observed simultaneously by more than one fluorescence
detector, the difference in the reconstructed shower parameters can be
used to infer the resolution of the $\Xmax$ reconstruction. It is
about 20~\gcm.
\subsection{Field of View Bias}
\begin{figure*}
  \subfigure[$\Xmax$ distribution (proton, {\scshape Sibyll}).]
  {\includegraphics[width=0.325\linewidth]
   {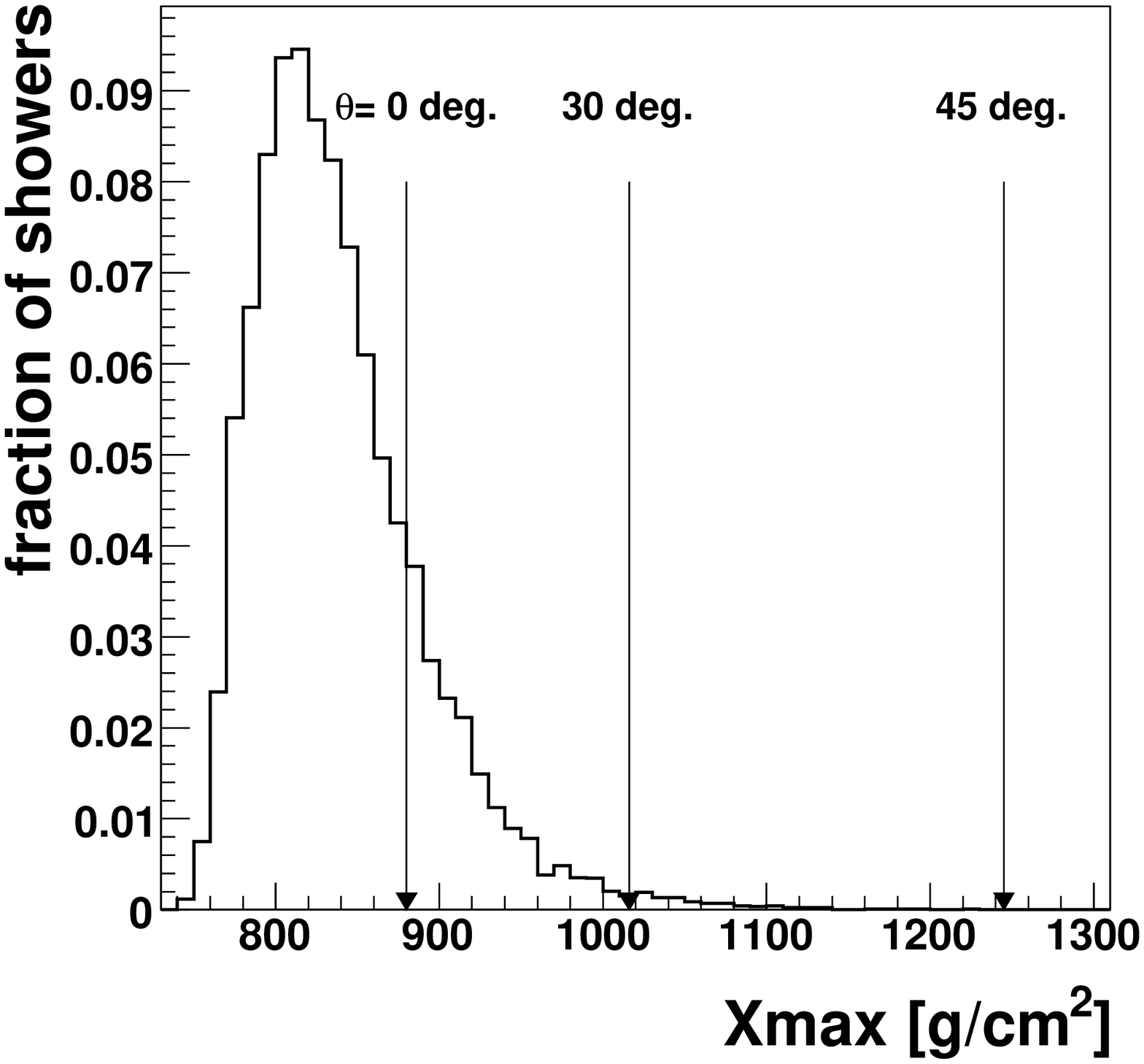}
  \label{fig_FOVFid0}}
  \subfigure[$\Xmax$ bias.]{\includegraphics[clip,bb=0 -20 567 539,
      width=0.325\linewidth]
   {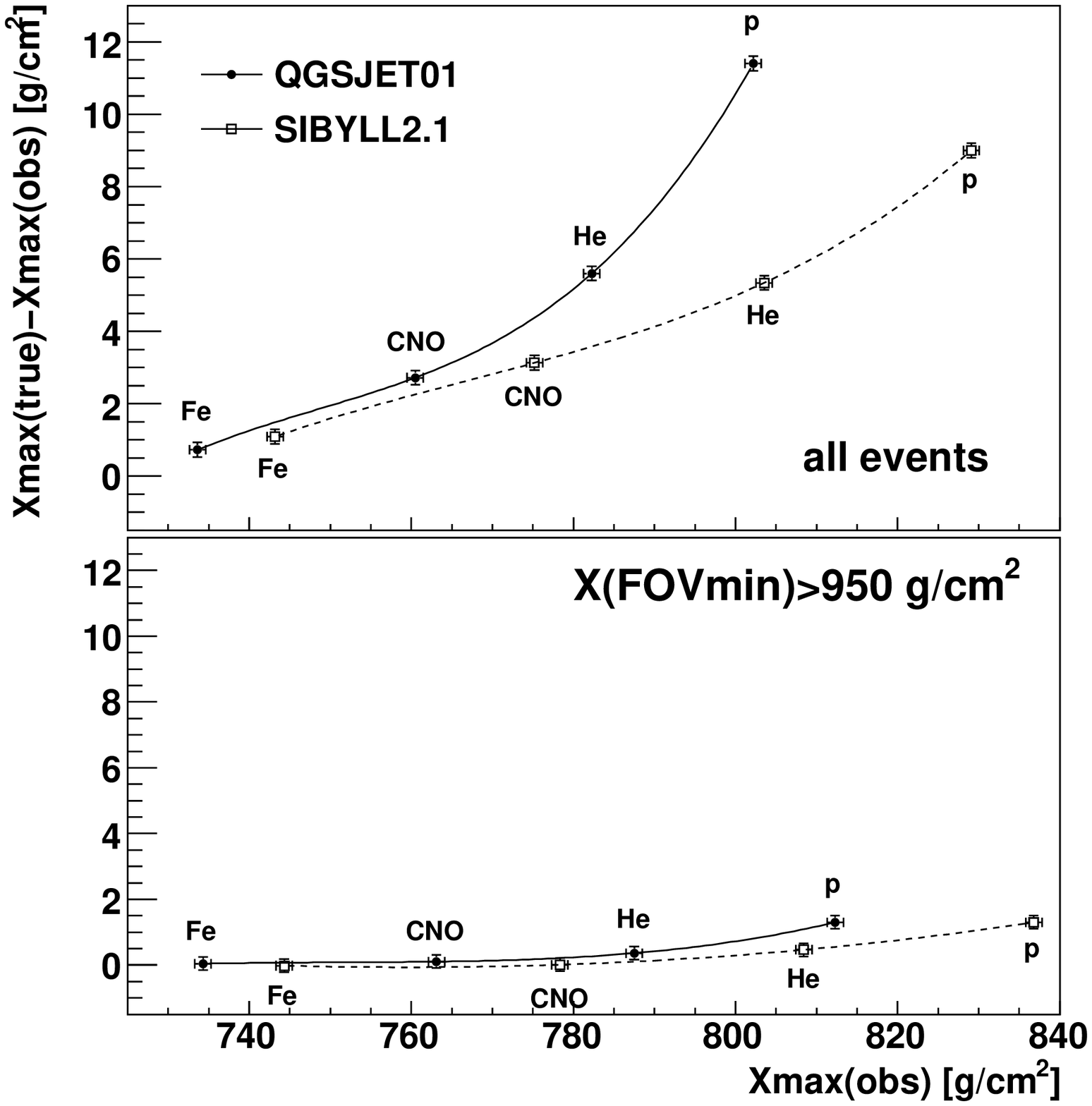}
  \label{fig_FOVFid1}}
  \subfigure[Fiducial volume cuts.]
            {\includegraphics[clip,bb=0 -80 567 384,width=0.335\linewidth]
   {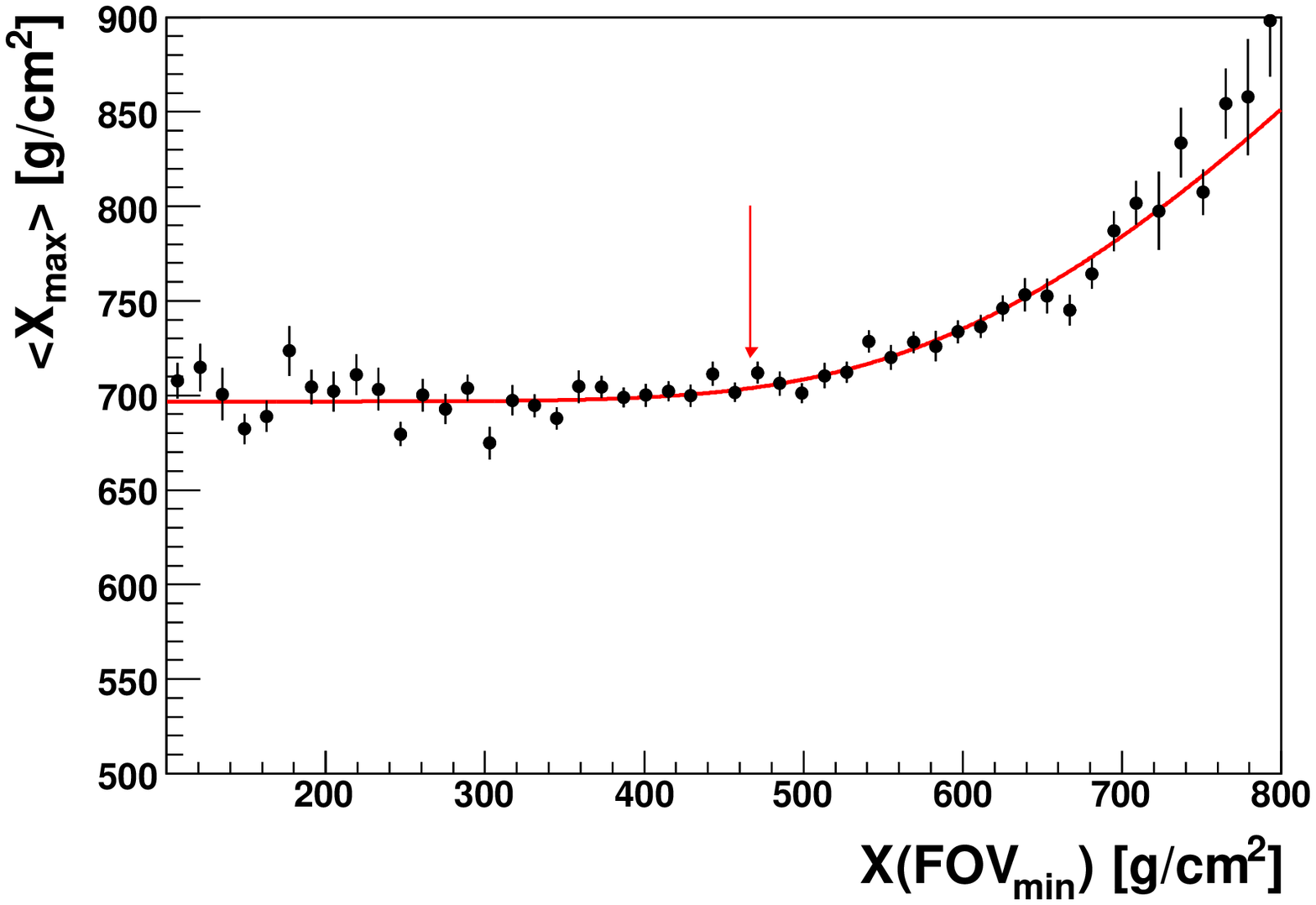}
  \label{fig_FOVFid2}}
  \caption{Field of view bias. Left and middle panel: Air shower simulations at 
           E=10$^{19.5}$ \eV. Right panel: Data at 10$^{18}$ \eV$<E<$10$^{18.25}$ \eV.}
  \label{fig_FOVFid}
  \vspace*{-0.3cm}
\end{figure*}
The energy and shower maximum can only be reliably measured, if
$\Xmax$ is in the field of view of the telescopes (otherwise only the
rising or falling edge of the profile is detected). As illustrated in
Fig.~\ref{fig_fov}, this may lead to a biased event selection, if
both, events with a small or large $\Xmax$, are not included in the
analysis because of their poor reconstruction.  The current Auger
fluorescence detectors cover an elevation range from
$\Omega_1=1.5^\circ$ to $\Omega_2=30^\circ$ and therefore the
observable heights for vertical tracks are between $R\tan
\Omega_1<h_v<R\tan \Omega_2$, where R denotes the distance of the
shower core to the fluorescence detector. That is, the farther away
from a fluorescence detector a track is detected, the smaller becomes
the observable upper slant depth boundary
$X(\mathrm{FOV}_\mathrm{min})$. Similarly the lower slant depth
boundary $X(\mathrm{FOV}_\mathrm{max})$ becomes larger for near
showers.\\ The absolute magnitude of the field of view bias depends of
course on the, a priori unknown, distribution of the depth of shower
maxima of cosmic ray showers.  It is instructive to study the order of
magnitude of the field of view bias for simulated showers with a known
composition. The expected $\Xmax$ distribution for protons with an
energy of 10$^{19.5}$ \eV is shown in Fig.~\ref{fig_FOVFid0}. The
upper viewable slant depth limits are indicated as arrows for
different zenith angles.  For vertical tracks it is around 880~\gcm at
the southern observatory, which is obviously not sufficient to detect
the full $\Xmax$ distribution. A convolution of the truncated $\Xmax$
distributions with the arrival directions of isotropic cosmic rays
($\mathrm{d}N/\mathrm{d}\cos\theta\varpropto\cos\theta$) leads to an
effective bias for the measurement of the mean of the distribution.
As can be seen in the upper panel of Fig.~\ref{fig_FOVFid1}, this bias
can be as large as 12~\gcm.  Due to the unknown composition and the
theoretical uncertainties of air shower simulations it can not be
corrected for a posteriori, since there is no unambiguous correction
given a measured $\meanXmax$.\\ The strategy followed in this analysis
is therefore to ignore showers that have a too small viewable
slant depth range. The rejection of showers with
$X(\mathrm{FOV}_\mathrm{min})<$950~\gcm almost completely removes the
bias on $\meanXmax$ in the simulation, cf. lower panel of
Fig.~\ref{fig_FOVFid1}.  The {\itshape fiducial volume cuts} that
ensure an unbiased $\meanXmax$ can be derived directly from data as
illustrated in Fig.~\ref{fig_FOVFid2}: The dependence of $\meanXmax$
on the slant depth boundaries is studied and events are only selected
in the region where $\meanXmax$ is constant.
\subsection{Results}
\label{ref:fdresults}
The $\meanXmax$ results presented here were derived from hybrid data
recorded between December 2004 and April 2007. In order to ensure a
good $\Xmax$ resolution, the following quality cuts were applied to
the event sample: The reconstructed $\Xmax$ should lie within the
observed shower profile and the reduced $\chi^2$ of a fit with a
Gaisser-Hillas function should not exceed 2.5. Moreover, insignificant
shower maxima are rejected by requiring that the $\chi^2$ of a linear
fit to the longitudinal profile exceeds the Gaisser-Hillas fit
$\chi^2$ by at least four. Finally, the estimated uncertainties of the
shower maximum and total energy must be smaller than 40~\gcm{} and
20\%, respectively. Moreover, the fiducial volume cuts explained in
the last section are applied.\\ The systematic uncertainties of the
atmospheric properties, event reconstruction and mass acceptance add
up to a total estimated systematic uncertainty of $\meanXmax$ 
that is $\lesssim$~15~\gcm{} at low energies and $\lesssim$ 11~\gcm{} above
\begin{figure}
 \hspace*{-0.5cm} \includegraphics[width=1.1\linewidth]
   {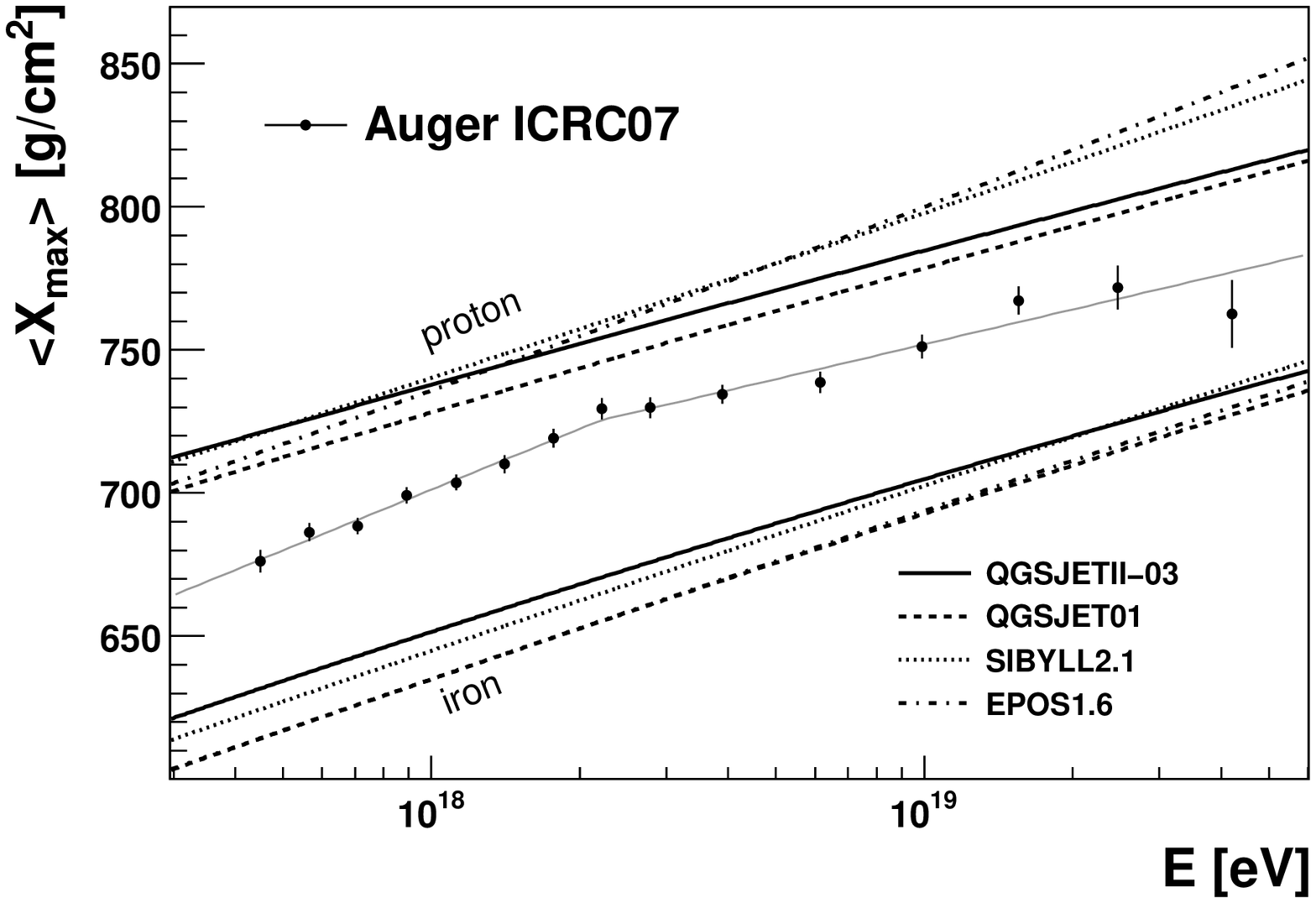}
  \caption[ER]{$\meanXmax$ as a function of energy~\cite{bib:icrcelong}.}
  \label{fig_ER}
  \vspace*{-0.3cm}
\end{figure}
10$^{18}$~\eV{}.\\ In Fig.~\ref{fig_ER} the mean $\Xmax$ as a function
of energy is shown along with predictions from air shower simulations
\cite{bib:showersim}.  As can be seen, the measurement favors a mixed
composition at all energies.

A simple linear fit, yields an
elongation rate of 54$\pm$2 (stat.)~\gcm{}/decade, but does not
describe our data very well ($\chi^2/$Ndf$=24/13$, P$<$3\%).
\begin{figure}
\hspace*{-0.5cm}  \includegraphics[width=1.1\linewidth]
   {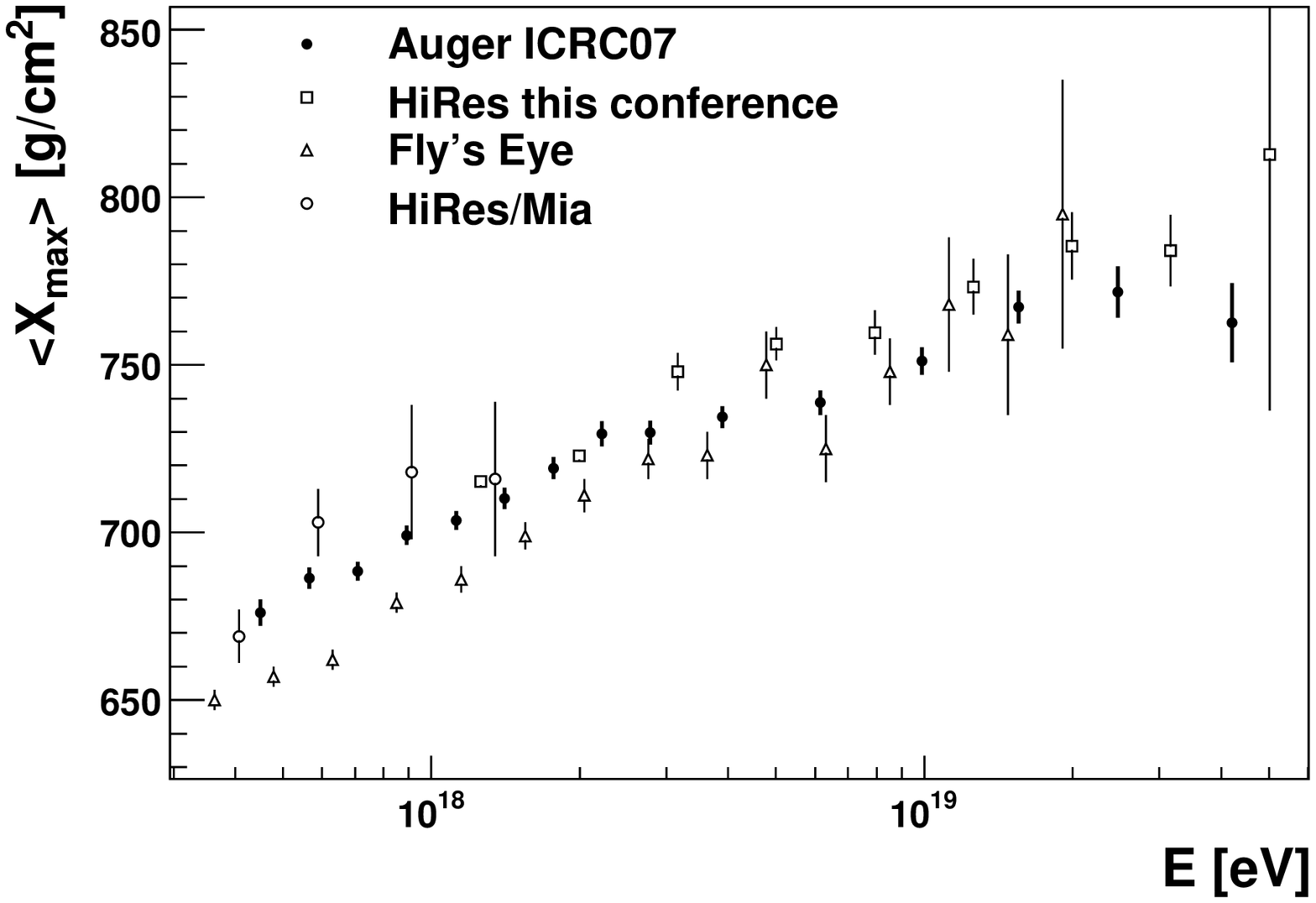}
  \caption[ER]{{Comparison to $\meanXmax$ from other experiments \cite{bib:oldXmax,bib:HiRes}.}}
  \label{fig_Others}
  \vspace*{-0.3cm}
\end{figure}
Allowing for a break in the elongation rate at an energy
$E_\mathrm{b}$ leads to a satisfactory fit with $\chi^2/$Ndf$=9/11$,
P$=$63\% and $D_{10}=71\pm5$ (stat.)~\gcm{}/decade below
$E_\mathrm{b}=10^{18.35}$~\eV{} and $D_{10}=40\pm4$
(stat.)~\gcm{}/decade above this energy. This fit is indicated as a
gray line in Fig.~\ref{fig_ER}.

Due to the uncertainties of hadronic interaction at highest energies,
the interpretation of these elongation rates is, however, ambiguous.
Using {\scshape QGSJetII} the data suggests a moderate lightening of the primary
cosmic ray composition at low energies and an almost
constant composition at high energies, whereas the {\scshape EPOS} proton
elongation rate is clearly larger than the measured one at high
energies, which would indicate a transition from light to heavy
elements. Theses ambiguities will be partially resolved by the
analysis of the $\Xmax$ fluctuations as an additional mass sensitive
parameter.

A comparison with measurements of other
experiments is presented in Fig.~\ref{fig_Others}. Taking into account
the individual systematic uncertainties of each experiment there is a
reasonable agreement of the $\meanXmax$ values. It is, however,
worthwhile noting that HiRes uses a different definition of $\Xmax$
(see~\cite{bib:HiRes}) that is about 10~\gcm shallower than the one
commonly used (depth of non-parametric shower maximum). If this is
taken into account, the elongation rates of Auger and HiRes show a
discrepancy in the $\Xmax$-scale of about 30~\gcm.
%
%
\section{Surface Detector Measurements}
\label{sec:sdcompo}
\begin{figure}
  \includegraphics[width=\linewidth]
   {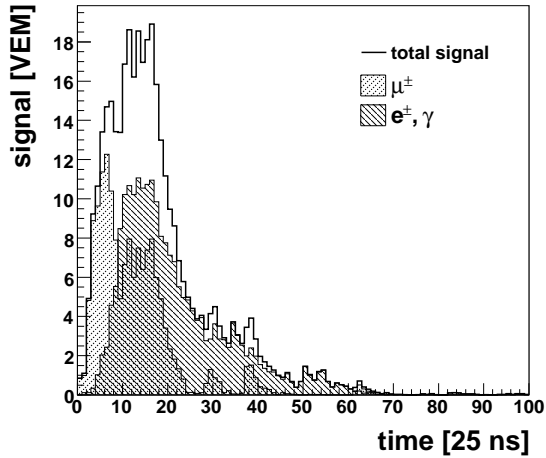}
  \caption{Simulated air shower signal in a water Cherenkov detector}
  \label{fig_VEM}
  \vspace*{-0.3cm}
\end{figure}
As we have explained in the last section, the longitudinal 
air shower development is a good tracer for the primary composition
that can be measured very precisely with fluorescence telescopes.
However, these telescopes can only take
data in moonless nights, whereas the surface detector has a duty cycle
of almost 100\% and correspondingly, the available event statistics
are about a factor ten higher. Because of this, and the obviously independent
systematic uncertainties, it is worthwhile to study ground level observables
that are sensitive to the primary composition as well.

Since the amount of electromagnetic particles at ground depends for
a given energy on the distance to the shower maximum it depends also
on the primary mass. 
Moreover, the number of ground level muons evolves differently 
with energy for different primary masses. From the Heitler model
one expects
\begin{equation}
N_\mu \varpropto E^\beta\,A^{1-\beta}
\end{equation}
where $\beta$ depends on the multiplicity and inelasticity of hadronic
interaction (and is thus, as the above $\alpha$, subject to theoretical
uncertainties).\\ Since the water Cherenkov detectors of the Pierre
Auger Observatory cannot explicitly discriminate between the
electromagnetic and muonic air shower components on an event by event
basis (but see Sec.~\ref{sec:outlook}), a variety of 'indirect'
experimental variables are currently investigated to relate the
surface detector data to the primary mass~\cite{Healy:2007sb}: The
signal rise time, its asymmetry and the signal shape
analysis\footnote{For yet another composition sensitive parameter, the
  shower front curvature, see~\cite{bib:curvature}.}.
\subsection{Risetime}
For each event, the water Cherenkov detectors record the signal as a
function of time (FADC traces). This provides a possibility to distinguish the
muonic and electromagnetic component, since the former travel in almost
straight lines through the atmosphere, whereas the latter undergo
multiple scattering on their way to ground. Therefore, the two
components have different path lengths and arrival times at ground.
This is illustrated in Fig.~\ref{fig_VEM}, where an example of a
simulation \cite{Heck:1998vt,Argiro:2007qg} of the surface detector
response is shown. The {\itshape rise time} $t_{1/2}$
\begin{figure}
  \centering
  \includegraphics[width=0.65\linewidth]
   {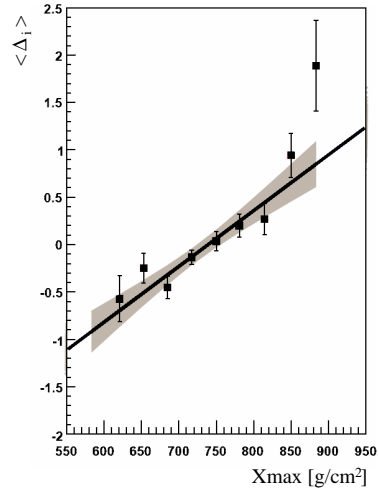}
  \caption{Correlation of rise time and $\Xmax$}
  \label{fig_RiseTime}
  \vspace*{-0.3cm}
\end{figure}
\cite{bib:risetime} is the time it takes until the cumulative signal
grows from 0.1 to 0.5. It is a measure of the muon to electron ratio
in a tank and depends on the primary mass, angle of incidence $\theta$
and distance to the shower core $r$.  The $r$ and $\theta$ dependence
is marginalized by a comparison to the average rise time $\langle
t_{1/2}(\theta,r)\rangle$ and the pull $\Delta_i=(t_{1/2}-\langle
t_{1/2}(\theta,r)\rangle)/\sigma(t_{1/2})$ is averaged over the
stations of an event to yield
$\langle\Delta_i\rangle$~\cite{bib:thesisBenSmith}.  This quantity
could in principle be compared to air shower simulations.  However, as
previous studies showed, the number of muons measured in Auger can not
be reproduced by these calculations~\cite{Engel:2007cm}. Therefore,
instead of a direct comparison of $\langle\Delta_i\rangle$ to air
shower simulations, it is related to the shower maximum. For this
purpose, the subset of events that have been detected by both, the
fluorescence and surface detector, are used to establish the relation
between $\langle\Delta_i\rangle$ and $\Xmax$. As can be seen in
Fig.~\ref{fig_RiseTime}, there is a linear correlation between the two
variables, with which the surface detector data can be calibrated to
yield $\Xmax(\langle\Delta_i\rangle)$, with which it is possible to
measure the elongation rate with surface detector data.
\begin{figure}
  \centering
  \includegraphics[width=0.88\linewidth]
   {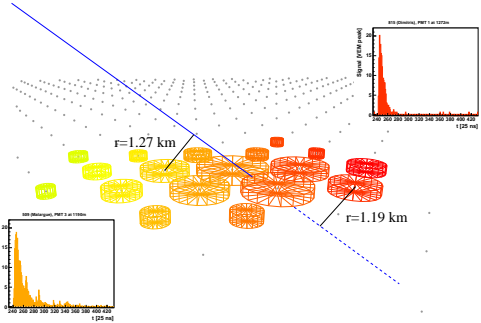}
  \caption{Asymetry of station signals ($\theta$=54$^\circ$, E=55$\pm$2 \EeV)}
  \label{fig_asymetry}
  \vspace*{-0.3cm}
\end{figure}
\begin{figure}
  \centering
  \includegraphics[width=0.765\linewidth]
   {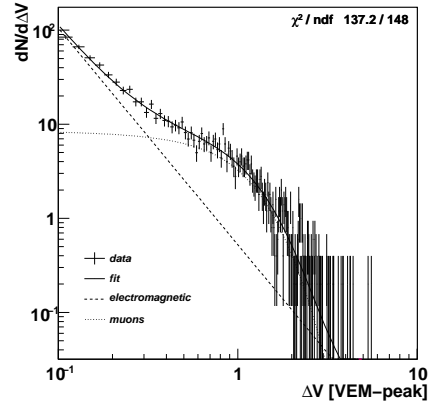}
  \caption{Distribution of jumps for events at a given energy, zenith
    angle and distance to the core range.}
  \label{fig_muons}
  \vspace*{-0.3cm}
\end{figure}
\subsection{Risetime Asymmetry}
The SD stations measure one stage of the shower development for near
vertical events. In case of inclined showers, however, considerably
different shower ages can be observed, depending on whether the
station is up- or downstream of the incoming shower direction.  This
is illustrated in Fig.~\ref{fig_asymetry}, where the signal traces of
two stations (both at the same shower plane distance $r$) are shown.
The upstream trace is rather broad, which is compatible with a large
fraction of electromagnetic particles.  The downstream trace is
narrower indicating that most of the electromagnetic component has
been attenuated and the signal is dominated by muons.  The
corresponding {\itshape asymmetry}~\cite{bib:riseasym} in the rise
time is thus a measure of the longitudinal development of the ratio of
electromagnetic to muonic particles.  The amplitude of the asymmetry
changes with the zenith angle, i.e.\ distance to the shower maximum.
The angle at which it reaches its maximum can be
used as an estimator for the primary composition.
\subsection{Signal Shape}
Since a muon deposits much more energy (typically 240 MeV) in a water
tank than an electron or photon (about 10 MeV), spikes are produced
over the smoother electromagnetic background in the FADC time
traces. Therefore, muons can be identified by searching for sudden
variations, in the signal from one FADC bin to the next 
(cf.\ Fig~\ref{fig_VEM}). The expected distributions of variations for
purely electromagnetic and muonic traces can be fitted to the measured
distribution as shown in Fig.~\ref{fig_muons}. In that way, the muonic
signal can be determined on a statistical basis with a resolution of
about 25\% and the number of muons as a function of energy can be
compared to predictions from air shower simulations to estimate the
primary composition.
%
%
\section{Conclusions and Outlook}
\label{sec:outlook}
In this article we presented the measurement of the
elongation rate from data collected with the fluorescence
telescopes of the Pierre Auger Observatory. When compared
to predictions from air shower simulation the $\meanXmax$
data favors a mixed composition at all energies. This measurement
will be soon updated with larger statistics and an analysis
of the fluctuations of the shower maximum. At lower energies
the systematic uncertainties will be reduced by additional
telescopes that cover the field of view from 30 
to 60 degree~\cite{Klages:2007zza}.\\ 
A variety of surface detector observables are sensitive
to the mass composition. These parameters
have a somewhat worse mass resolution than $\Xmax$, which is, however,
outweighed by the large event statistics collected by the ground array.
Due to the hybrid design of the Pierre Auger Observatory, the correlation
between the lateral and longitudinal shower parameters can be studied 
and additional muon detectors will soon allow for an event-by-event
measurement of the muon content of a shower~\cite{Etchegoyen:2007ai}.
%
%

%
%
\end{document}